\documentclass[prb,twocolumn,showpacs,amsmath,amssymb,superscriptaddress,floatfix]{revtex4}

\usepackage[english]{babel}
\usepackage{amsfonts}
\usepackage{graphicx}
\usepackage{times}

\begin{document}

\title{ Non-equilibrium dynamics of the Holstein polaron driven by an external electric field
}

\author{Lev Vidmar}
\affiliation{J. Stefan Institute, 1000 Ljubljana, Slovenia}

\author{Janez \surname{Bon\v ca}}
\affiliation{J. Stefan Institute, 1000 Ljubljana, Slovenia}
\affiliation{Faculty of Mathematics and Physics, University of Ljubljana, 1000
Ljubljana, Slovenia}

\author{Marcin \surname{Mierzejewski}}
\affiliation{Institute of Physics, University of Silesia, 40-007 Katowice, Poland}
\affiliation{J. Stefan Institute, 1000 Ljubljana, Slovenia}

\author{Peter \surname{Prelov\v sek}}
\affiliation{J. Stefan Institute, 1000 Ljubljana, Slovenia}
\affiliation{Faculty of Mathematics and Physics, University of Ljubljana, 1000
Ljubljana, Slovenia}

\author{Stuart A. Trugman}
\affiliation{Theoretical Division, Los Alamos National Laboratory, Los Alamos, New Mexico 87545, USA}




\date{\today}
\begin{abstract}
This work represents  a fundamental  study  of a  Holstein polaron  in one dimension driven  away from the ground state  by a constant electric field. 
Taking fully  into account quantum effects we follow the time-evolution  of the system from its ground state as the constant electric field is switched on at $t=0$, until it reaches a steady state. 
At weak  electron phonon coupling (EP) the system experiences damped Bloch oscillations (BO) characteristic for  noninteracting  electron band. An analytic expression of the steady state current is proposed  in terms of weak EP coupling and large electric field.
For moderate values of EP coupling the oscillations are almost  critically  damped and the system reaches the steady state after a short time.
In the strong coupling limit  weakly damped  BO, consistent with nearly  adiabatic evolution within  the polaron band, persist up to extremely large  electric fields.
A traveling polaron under the influence of the electric field leaves behind a trail of phonon excitations absorbing the excess energy gained from the electric field.  The shape of the traveling polaron is investigated in details. 
\end{abstract}  

 \maketitle


\section{Introduction}
Research in the field of   non-equilibrium dynamics of complex quantum systems constitutes a formidable theoretical challenge. Many advanced numerical techniques, ranging from exact-diagonalization \cite{oka}, expansion using Chebyshev polynomials \cite{fehske4},  time-dependent density matrix renormalization group \cite{white}, to non-equilibrium dynamical mean field techniques \cite{freericks} have been developed to tackle this complex problem.

%

More than forty  years ago using a path-integral approach Thornber and Feynman\cite{thornber} discovered  that an electron in a parabolic band, driven by the  electric field, acquires a constant velocity due to emission of phonons.  
Later approaches to polaron motion in high electric field used  Boltzmann equations \cite{khan},  the high-field drift velocity was estimated via phonon-assisted hopping between different rungs of Wannier-Stark states using  rate equations \cite{emin,rott}. In ref.~\cite{bonca5} the  one-dimensional Holstein polaron problem in strong electric field has been mapped on a nonstandard Bethe lattice. It has been realized  that keeping full quantum coherence between many-body states is crucial to obtain finite drift velocity for dispersionless optical phonons.  Extensive research of polaron dynamics has been conducted within the semiclassical Su-Shrieffer-Heeger model to describe properties of conjugated polymers that may be used in a variety of applications like molecular electronics or light-emitting diodes \cite{johan1,basko,johan2,LiuX,LiY,QiuY}. 
Polaron formation and its influence on transport properties has also been investigated in the context of DNA molecules within the semiclassical Peyrard-Bishop-Holstein model \cite{Peyrard,Komineas,Maniadis1,Maniadis2,Berashevich,Diaz1,Diaz2}
and other polaron-like models \cite{Chang,Yamada,Macia}.

Bloch oscillations (BO) represent a fundamental phenomenon in quantum mechanics where a charged particle in a periodic potential exhibits a periodic motion when exposed to an uniform external electric field.
Since the electrons in solids can dissipate energy due to scattering from  inelastic degrees of freedom on a time scale usually shorter than a typical Bloch time $t_B$,
it took a long time until the first experimental observation of BO was carried out on semiconducting superlattices
\cite{Feldmann92,Leo92,Waschke93,Mendez88,Voisin88} and later in optical potentials \cite{Bendahan96,Wilkinson96,Niu96}.
Nowadays the concept of BO is frequently present in a variety of different fields, for instance atomic Bose-Einstein condensates in optical lattices \cite{Anderson98,Morsch01,Ferrari06,Kolovsky10}, interacting quantum few-body systems \cite{Buchleitner03,Dias07,Khomeriki10} or organic molecules \cite{Dominguez03,LiY,Diaz1,Diaz2}.
However, the description of damping of BO in dissipative medium remains a challenging task.

By choosing the Holstein Hamiltonian as one of the simplest model systems describing the interaction between a fermion and phonons, we are able to investigate the field-induced acceleration of the polaron, which simultaneously dissipates the energy by inelastic scattering on optical phonons while maintaining the full quantum nature of the problem.
Following the time evolution of the ground state when the electric field is switched on at time  $t=0$, we show how the polaron reaches the steady state and consequently develops a constant non-zero velocity.
In particular, we calculate the steady-state current vs. voltage characteristics of the Holstein polaron for different regimes of electron-phonon couplings.

We discuss the Holstein model in one spacial dimension and give a brief overview of the numerical method in the second paragraph. In the third we discuss numerical results. Here we give special emphasis on the time evolution from the ground state towards the steady state by presenting various correlation functions in different EP coupling regimes. We compare our results with a simple Landau Zener model and  follow the time evolution of the polaron as it starts propagating after switching on the electric field. As a focal point of this work we  discuss the dependence of  the steady state current on the  EP coupling and electric field.  In the last paragraph we give conclusions.

\section{Model and numerical method}


We analyze the one-dimensional Holstein model with a single electron, threaded by a time-dependent flux:
\begin{eqnarray} \label{hol}
\vspace*{-0.0cm}
H &=& -t_0\sum_{i}(e^{i \theta(t)}c^\dagger_{i} c_{i+1} +\mathrm{H.c.})\nonumber \\
 &+& {g} \sum_{i} n_{i} (a_{i}^\dagger + a_{i}) +
\omega_0\sum_{i} a_{i}^\dagger  a_{i}, \label{ham}
\end{eqnarray}
where $c^\dagger_{i}$ and $a^\dagger_{i}$ are electron and phonon creation operators at site $i$, respectively, and $n_{i} = c^\dagger_{i}c_{i}$ is electron density. 
$\omega_0$ denotes a dispersionless optical phonon frequency and $t_0$ nearest-neighbor hopping amplitude.
The dimensionless EP coupling strength is $\lambda = g^2/2t_0\omega_0$. The constant electric field $F$ that is switched on at time $t=0$  enters the Hamiltonian in Eq.~\ref{hol} through the time-dependent phase $\theta(t)=-Ft$ for $t\geq 0$. We measure electric field $F$ in units of $[t_0/e_0a]$ where $e_0$ is the unit charge and $a$ is the lattice distance. We furthermore measure time in units of $[\hbar/t_0]$. Unless otherwise specified, from here on we set $a=e_0=\hbar=t_0=1$. 

To solve the time-dependent Hamiltonian for  a single electron coupled to phonon degrees of freedom we use  an  improved numerical method, originally introduced in Ref.~\cite{bonca1}, that led to numerically exact solutions of the polaron ground and low-lying  excited state properties. The method constructs the variational Hilbert space (VHS) starting from the single-electron Bloch state $c_{\bf k}^{\dagger} \vert \emptyset{\rangle}$ with no phonons  on an infinite lattice. The VHS is then generated by applying the off-diagonal terms of Hamiltonian (\ref{hol})
\begin{equation}
\left \{ \vert \phi_{{\bf k},l}^{(N_h,M)}{\rangle}\right \} = \left ( H_{\rm kin} + 
H_{\rm g}^M\right )^{N_h} c_{\bf k}^{\dagger} \vert \emptyset{\rangle},
\label{gen}
\end{equation}
where $H_{\rm kin}$ and $H_{\rm g}$ correspond to the first and the second term of the Hamiltonian in Eq.~\ref{hol}, respectively. Parameters $N_h$ and $M$ determine the size of the VHS.  In addition, $N_h-1$ represents  the maximum distance between the electron and the phonon quanta and $N_h*M$ is the maximum number of phonon quanta contained in the Hilbert space. The parameter $M>1$  (Ref.~\cite{bonca3})  ensures good convergence in  the strong EP coupling regime that contains  multiple phonon excitations.
To reach weak coupling regime, $\lambda<<1$, we introduce an additional parameter $N_{\mathrm{phmax}}$ limiting the maximum number of phonon quanta, which enables construction of VHS as large as $N_{h}= 40$.


We first solve the Hamiltonian in Eq.~\ref{hol} for $F=0$, i.e., we calculate the (zero temperature) 
polaron ground state 
\cite{bonca1,wellein97,ciuchi97,jeckelmann98,romero98,alexandrov07,berciu07b,osor3,marsiglio10,alvermann10,Zoli10}.
Then we switch on the uniform electric field and start the time propagation from the initial state using the time-dependent Lanczos technique \cite{park}.
We manage to find numerically accurate results of the model away from equilibrium while maintaining full quantum description of phonons.
Since we are dealing with a single particle in an infinite system, we compute time-dependent average of the current operator  $j(t)=\langle \hat I(t)\rangle$, where
\begin{equation}
\hat I(t) = i \left( \sum_{l}e^{-iFt }c^\dagger_{l} c_{l+1} - \mathrm{H.c.} \right) .
\end{equation}
In the case of a time-independent  field $F$, the time-integral of the current is directly related to a change of the total energy 
\begin{equation}
\int_0^tj(t')dt'=\Delta h(t)/F= x(t),
\label{sumrule}
\end{equation}
where $\Delta h(t) = \langle H(t)\rangle - \langle H(t=0)\rangle$ and $x(t)$ represents the travelled distance \cite{marcin}. 

\section{Numerical results}

\subsection{Time evolution from the ground state towards the steady state}

\begin{figure}[!tbh]
\includegraphics[width=1.0\columnwidth]{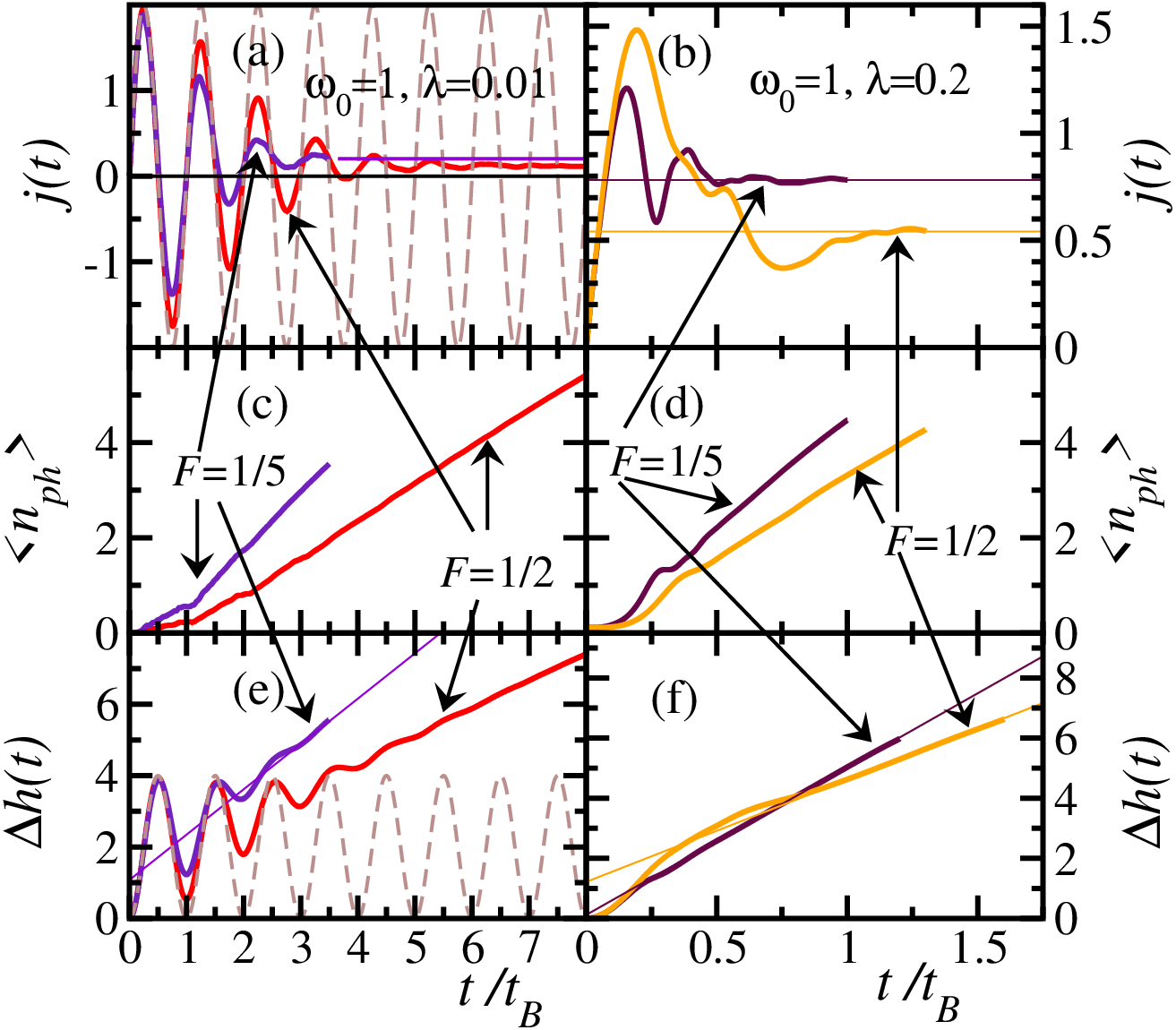}
\caption{(Color online) $j(t)$ vs. $t/t_B$ for two values of $F=1/5$ and $1/2$ for $\omega_0=1$,  and two distinct values of $\lambda$: (a) $\lambda=0.01$ and (b) $\lambda=0.2$. Thin dashed line in (a) represents $j(t)$ for $\lambda=0$, thin horizontal lines in (a) and (b) indicate steady-state  values $\bar j$; $\langle n_{ph}\rangle$ is shown in (c) and (d) for the same set of parameters as in (a) and (b), respectively; corresponding averages $\Delta h(t)$ are displayed in (e) and (f).
Thin dashed line in (e) represents $\Delta h(t)$ for $\lambda=0$.
The accuracy of time propagation was checked by comparison of the energy-gain sum rule, Eq.~\ref{sumrule}. Parameters, defining functional generator (Eq.~\ref{gen}) were $N_h=40$, $M=1$, and $N_{\mathrm{phmax}}=6$ for $F=1/5$ and $N_h=28$, $M=1$ and $N_{\mathrm{phmax}}=8$ for $F=1/2$. In this and all subsequent figures we used up to $N_{\mathrm{st}}\sim 15 \times 10^6$ states in the Hilbert space and $N_{\mathrm{step}}=2000$ time steps within  each $t_B$.
Different sizes of VHS were used  to check the convergence in the thermodynamic limit. 
Thin straight lines represent $t\to \infty$ extrapolations.
}\label{fig1}
\end{figure}

We first present results obtained near the noninteracting  limit,  {\it i.e.} at $\lambda=0.01$, where $j(t)$ displays damped BO around $j(t\to \infty)>0$, see Fig.~\ref{fig1}(a). The period and the initial amplitude of BO at small $t$ are consistent with BO of a free electron, denoted with thin dashed line in Fig.~\ref{fig1}(a). Damping is due to inelastic scattering on phonons that is in turn reflected in a monotonic increase of the average phonon number $\langle n_{ph}\rangle$ with time, as depicted in Fig.~\ref{fig1}(c). 
Damping is, however, not the most important consequence of inelastic scattering. Notably,  $j(t)$ approaches a positive  steady state current $\bar j$ for $t>t_s\sim  4t_B$ and $6t_B$  at $F=1/5$ and $F=1/2$, respectively,  where $t_B$ denotes the   Bloch oscillation period $t_B=2\pi/F$. 
Note that $\bar j<< j_{\mathrm{max}}=2$. 
The dependence of the steady state current $\bar j$ on $F$ will be discussed further in the text. The steady state current as well emerges  as  a linear dependence  of the total energy on  time: $\Delta h(t)= F \bar j t+ \Delta h_0$, see  results in  Fig.~\ref{fig1}(e), where with increasing $t$, $\Delta h(t)$  approaches a straight line. 
In the steady state  we as well observe a linear increase of  $\langle n_{ph}\rangle$ vs. $t$. When comparing $\Delta h(t)$ and $\langle n_{ph}\rangle$ in the linear regime we find that $\Delta \dot{h}(t) = \omega_0 {\mathrm d}\langle n_{ph}\rangle/{\mathrm d}t$. This equality confirms an intuitive expectation   that  in the steady state  the total energy gain is entirely absorbed by the lattice.

On a more technical side we note that to reach a steady state, the Hilbert space used in our  calculation must contain  large enough set of excited states that in turn represent the reservoir for the absorption  of energy. 
For this reason, different Hilbert spaces were used, depending on the strength of EP coupling and the size of $F$, see as well Caption of Fig.~\ref{fig1}. 

At a larger value of EP coupling,  $\lambda=0.2$,  a somewhat   different physical picture emerges, 
as shown in Figs.~\ref{fig1}(b,d) and (f). The main differences can be summarized as: (i) BO become overdamped, (ii) $j(t)$ remains positive at all $t$, and (iii) $j(t)$ reaches a steady state after a short time $t_s\lesssim t_B$. Characteristic for a steady state are linear $t$ dependencies of $\langle n_{ph}\rangle$ and $\Delta h(t)$ in Figs.~\ref{fig1}(d) and (f), respectively. Common to all cases presented in Fig.~\ref{fig1}, is the emergence of a constant steady-state current for $t>t_s$. 

\begin{figure}[!tbh]
\includegraphics[width=1.0\columnwidth]{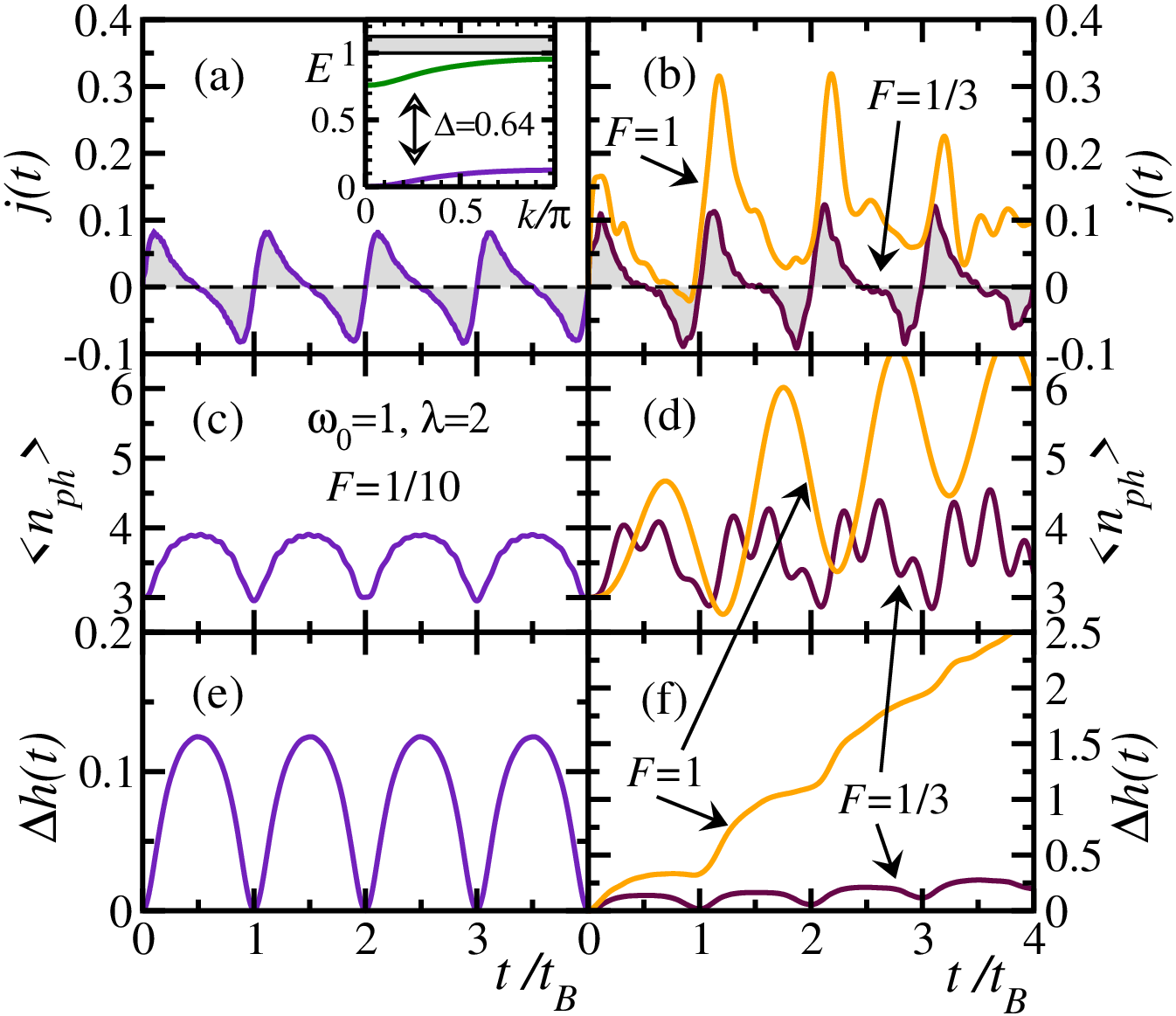}
\caption{(Color online)  $j(t)$ in (a) and (b),  $\langle n_{ph}\rangle$ in (c) and (d), and $\Delta h(t)$ in (e) and (f) vs. $t/t_B$ for $\omega_0=1$ and $\lambda=2$ and three different values of $F$ as indicated in figures. Inset in (a) shows the polaron spectrum (ground state, first excited state energy and the continuum vs. the wavevector $k$) for $\omega_0=1$ and $\lambda=2$. Note that there is a different vertical scale used in (e) and (f). 
We used $N_h=N_{\mathrm{phmax}}=20$ and $M=1$ with $N_{\mathrm{st}}=3\times 10^6$. 
}\label{fig2}
\end{figure}

In Fig.~\ref{fig2}(a) and (b) we present current vs. time in the strong coupling regime, {\it i.e.} at $\lambda=2.0$. At $F=1/10$ (Fig.~\ref{fig2}(a)) we observe nearly undamped  BO as the polaron adiabatically follows the polaron band. Regular oscillations in $\langle n_{ph}\rangle$ and $\Delta h(t)$ in Figs.~\ref{fig2}(c) and (e) portray  polaron averages nearly identical to their  ground state values  at corresponding  wavevectors $k=Ft=2\pi t/t_B$. The response of the system to external field is nearly elastic, since $\Delta h(t=l*t_B)\sim 0$ for any integer value of $l$. The average current remains  indistinguishable from zero in   the largest time interval tested with our calculation, {\it i.e.} $t \leq 20 t_B$. 
 
In order to illuminate this behavior we note that in  the strong coupling limit a large gap $\Delta$ exists  in the polaron excitation spectrum being  of the order of $\omega_0$.  The low-energy polaron excitation spectrum is presented for  $\omega_0=1$ and $\lambda=2$ in the inset of Fig.~\ref{fig2}(a)  where a gap $\Delta\sim 0.64$   separates the polaron band from  the excited polaron band \cite{bonca1,osor1,osor2,vidmar10a}, located just below the  continuum denoted by the grey area. 
At small $F<<\Delta$ there exist exponentially small probability for a nonadiabatic transition from the polaron band to the excited polaron band or/and into the continuum. 
 
For sufficiently large $F$ (see Figs.~\ref{fig2}(b,d,f) for $F=1/3$ and $1$), BO in $j(t)$ lose periodicity even though remnants of BO remain clearly visible, and the time averaged current becomes finite (nonzero).
Additional  frequencies appear in $\langle n_{ph}\rangle$ that indicate multiple phonon excitations due to polaron transitions to excited polaron bands. Moreover, the average value of $\langle n_{ph}\rangle$ between successive $t_B$ intervals increases. The total energy $\Delta h(t)$ as well increases in time. At large field, $F=1$, $\Delta h(t)$ approaches a straight line signaling the onset of a steady state.

\subsection{Determination of the threshold electric field using the Landau-Zener formalism}

The observed behavior in the strong coupling regime due to a large gap in the spectrum  resembles Landau-Zener  (LZ)  transition \cite{landau,zener},  where the probability for tunneling between bands in a two level system,  
\begin{equation}
H(t)=\left( \begin{matrix}
vF t &  \Delta/2\\
\Delta/2  &   -vF t   
\end{matrix}\right),\label{LZ}
\end{equation}
is given by
\begin{equation}
P=\exp\left[ -\pi{(\Delta/2)^2\over vF}\right],
\label{lz}
\end{equation}
where $\Delta$ is the energy gap between the two levels and $v$ is the velocity. Using Eq.~\ref{lz} we estimate the threshold electric field $F_{\mathrm {th}}$ using
\begin{equation}
F_{\mathrm {th}}={\left(\Delta/2\right)^2\over v}.
\label{fth}
\end{equation}
Such an  estimate has been used to determine the dielectric breakdown of the insulating half-filled Hubbard model~\cite{oka,oka1}. Applying Eq.~\ref{fth} to the specific case of $\lambda=2$ presented in  Fig.~\ref{fig2}, using $\Delta \sim 0.64$ and $v=j_{\mathrm{max}}\sim 0.1$, we obtain  $F_{\mathrm {th}}\sim 1.0$. LZ formalism gives roughly the correct order of magnitude of $F_{\mathrm {th}}$, since a noticeable current appears around $F=1/3$, as seen in Fig.~\ref{fig2}(b) and (f) as well as from steady state current, presented in Fig.~\ref{fig4}(c). 
One should however be mindful when considering the  LZ formalism. In the polaron case the band structure deviates significantly from the ideal LZ model with two hyperbolic bands. In the realistic case, multiple transitions occur from the polaron  band to a continuum of excited states composed from a polaron and additional phonon degrees of freedom.

\begin{figure}[!tbh]
\includegraphics[width=1.0\columnwidth]{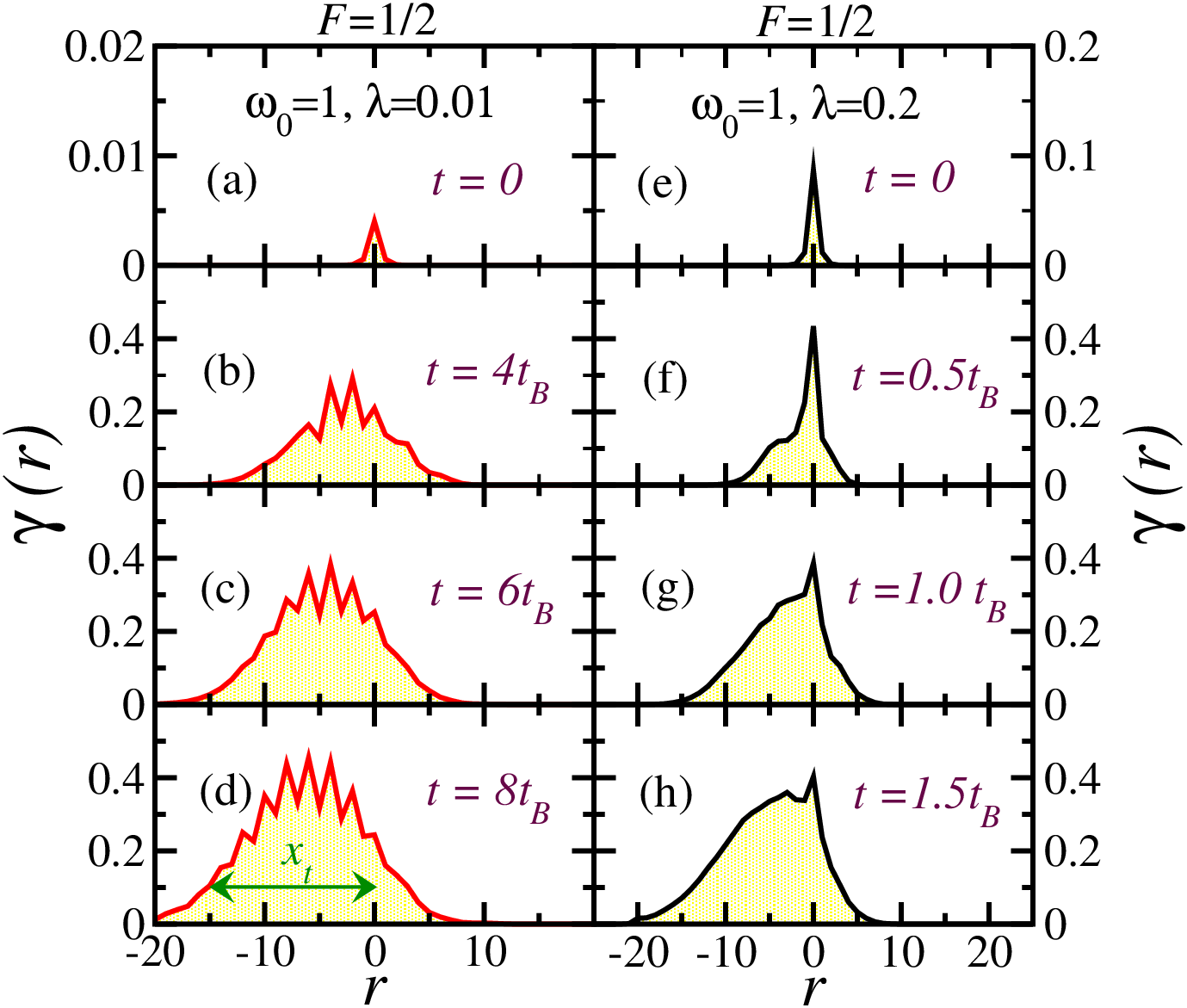}
\caption{(Color online) $\gamma(r)$ for $\omega_0=1$,  $\lambda=0.01$ and $F=1/2$  in (a) through (d) and $\lambda=0.2$ in (e) through (h) computed at different times. The electric field is switched on at $t=0$.  Note a vertical scale change between (a) and (b) as well as between (e) and (f). $x_t$ in (d) represents the travelled distance, see discussion in the text. 
}\label{fig3}
\end{figure}

\subsection{Time evolution of polaron}

In Fig.~\ref{fig3} we follow the time evolution of the polaron towards the steady state at $F=1/2$.
We compute the average number of phonon quanta located at a given distance $r$ from the electron 
\begin{equation}
\gamma(r) = \langle  \sum_i n_i a^+_{i+r}a_{i+r}\rangle,  
\label{gamma}
\end{equation}
fulfilling the  following  sum-rule   $\langle n_{ph}\rangle = \sum_r \gamma(r)$. At $t=0$, $\gamma(r)$ displays a pronounced peak at the position of the electron, {\it i.e.} at $r=0$, consistent with  the shape of the polaron in its $k=0$ ground state. After the electric field is switched on,  $\gamma(r)$ experiences  a compelling    time evolution with  three outstanding  characteristics: (i) the overall increase of $\gamma(r)$ with time, 
(ii) development of pronounced asymmetry of $\gamma(r)$ with respect to the electron position at $r=0$, and (iii) increased amount of polaron excitations in the forward direction. The overall increase of $\gamma(r)$ is consistent with the absorption of energy that is deposited  in increasing number of phonon excitations.  The asymmetry is a result of a growing phonon tail, extending  behind the moving polaron. Note that the polaron is moving from left to  right. In the long time limit, $\gamma$ is expected  to be approximately constant,
independent of $r$ and $t$, for r sufficiently negative. The average height of the polaron tail $\bar \gamma$ is due to 
energy conservation requirement independent of $\lambda$: 
\begin{equation}
x(t)F\sim \langle n_{ph}\rangle\omega_0\sim x(t)\bar \gamma \omega_0, 
\end{equation}
therefore $\bar \gamma\sim F/\omega_0$, compare  Figs.~\ref{fig3}(d) and (h). Note that this relation holds only when the system has reached a steady state.  The length  of the polaron tail is given by  the expression for the travelled distance $x(t)=\Delta h(t)/F$. At  $t=8t_B$ and $\lambda=0.01$ we obtain from Fig.~\ref{fig1}(e) $x_t=x(8t_B)\sim 14.8$, that fits well with the length of the  phonon tail in  Fig.~\ref{fig3}(d).

Rather  unexpected is  the pronounced increase of phonon excitations in the forward direction  where $\gamma(r) \gtrsim 0$ up to  $r\leq r_f\sim 5-7$ for all $t>0$  presented in  Fig.~\ref{fig3}. 
Since time evolution starts from the ground state at zero temperature, there are no phonon excitations present far ahead from the moving electron. A substantial  forward tail of phonon excitations is a consequence of damped BO. Indeed, $r_f$ compares well with  the Stark localization length, {\it i.e.}  $r_f\sim L_S=4/F=8$.  Yet another intriguing feature in $\gamma(r)$ emerges as regular oscillations in the polaron tail with a period $K=\omega_0/F=2$, clearly seen in the small $\lambda=0.01$ limit, see Figs.~\ref{fig3}(b) through  (d). 
At larger $\lambda=0.2$ these oscillations become overdamped. 

\subsection{Steady state current} 
 
\begin{figure}[!tbh]
\includegraphics[width=1.0\columnwidth]{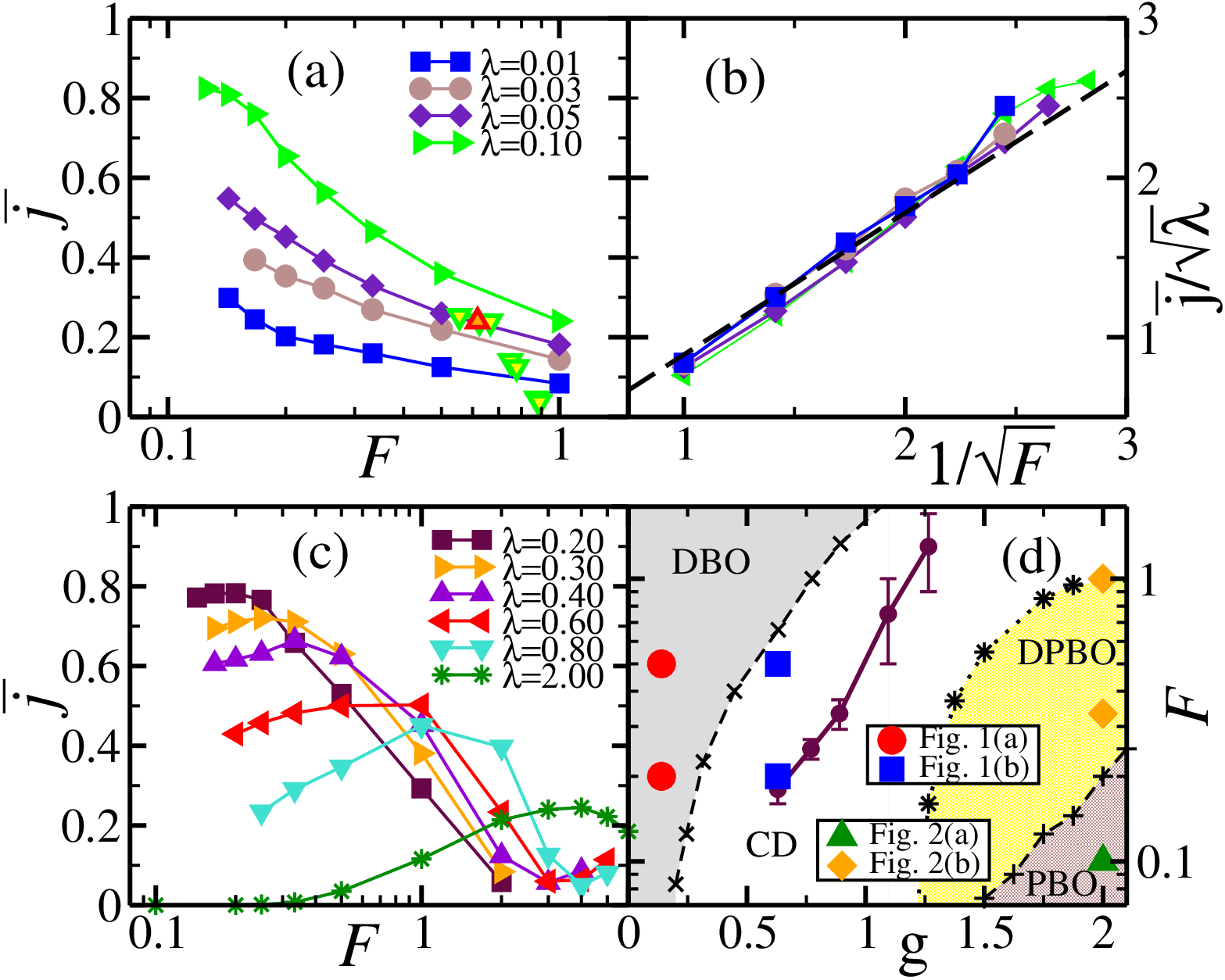}
\caption{(Color online) Steady state current $\bar j$ vs. $F$ in the weak coupling limit in (a) and in the weak to intermediate coupling regime in (c), scaling $\bar j/\sqrt{\lambda}$ vs. $1/\sqrt{F}$  in the weak coupling limit in (b), and  diagram, presenting different regimes, as described in the text (d). Also in (d) circles with error bars indicate positions in the diagram where for a fixed $g$, a maximum value $\bar j_{\mathrm{max}}$ was reached; isolated circles, squares and diamonds indicate values used for  Figs.\ref{fig1} and \ref{fig2}. Disconnected triangles (seven down  and one up) in (a)  represent $\bar j$ using non-integer values (seven rational and one irrational) of $1/F$, {\it i.e.} $F=5/9, 2/(1+\sqrt 5), 5/8,6/9,6/8,7/9, 7/8$ and $8/9$ at $\lambda=0.1$.   Different values of $N_h$, $M$ and $N_{\mathrm{phmax}}$ were used to ensure that 
error bars, where not specified, are smaller than sizes of the symbols. 
}\label{fig4}
\end{figure}

The focal point of this work is the calculation of the steady state current $\bar j$ and analysis  of its dependence on $F$ and $\lambda$. In Fig.~\ref{fig4} we present the current-voltage characteristics, i.e. $\bar j$ vs. $F$ for different values of $\lambda$. Note that the upper limit of $\bar j$ is given by the current amplitude  $j_{\mathrm max}=2$ in the  noninteracting system. We have  limited our calculations to commensurate values
 $F = \omega_0/ K$ with $K$ being integer, with two  exceptions: (i)  large $F>\omega_0$ where we have chosen $F=2\omega_0, 3\omega_0, \dots$ and (ii)  results presented with disconnected  triangles   in Fig.~\ref{fig4}(a), with details given in the figure caption. In the  regime $\lambda\leq 0.1$, presented in Fig.~\ref{fig4}(a), $\bar j$ decreases with increasing $F$ for $F\gtrsim 0.1$.  Our method does not yield steady state results  in the regime  $F\lesssim 0.1$ due to large Stark localization length  $L_S=4 t_0/F$. Since $\bar j=0$ for $F=0$ as well as in the opposite limit, when $F\to \infty$,  there must exist  a global maximum value $\bar j_{\mathrm{max}}$ that depends on $\lambda$. For  $\lambda=0.1$, $\bar j_{\mathrm{max}}\sim 0.82$, while for $\lambda< 0.1$, $\bar j_{\mathrm{max}}$ is reached somewhere  in the interval $0<F<0.1$, not accessible by the present numerical method. Choosing rational or even irrational values of $\omega_0/F$ leads to a decrease of $\bar j$ that nevertheless remains non-zero even in the latter case.  A sweep over continuous values of $F$ would lead to spikes in $\bar j$ located at integer values of $\omega_0/F$, as consistent with observations in previous works \cite{emin,rott,bonca5,Zhang02}. 


To gain further insight into the decrease of $\bar j$ with $F$, we plot in Fig.~\ref{fig4}(b) $\bar j/\sqrt{\lambda}$ vs. $1/\sqrt{F}$ and realize  that  curves approximately collapse onto a straight line. 
The revealed scaling with $1/\sqrt{F}$ is a clear signature that we are dealing with a coherent propagation between Stark 
states with identical total energy  that are spaced by $K=\omega_0/F$. This is in contrast with
the assumption of an incoherent hopping  between localized states \cite{emin,rott} , which would predict a 
dependence $j \propto 1/F$.  In turn our derivation, as presented in the Appendix, leads for integer $K>1$ as well as
for $\omega_0< W = 4t_0$  to a scaling of the maximum steady current 
\begin{equation}
j_{\mathrm 0} = \alpha\sqrt{\lambda \omega_0^3\over F}.
\label{jmax}
\end{equation}
While $j_{\mathrm{0}}$ cannot be directly compared to the average current $\bar j$, the functional dependence on $\lambda$ and $F$ is in good agreement with scaling in Fig.~\ref{fig4}(b) that leads to $\alpha\sim 0.89$ (fit is represented  by a dashed line).
The expression in Eq.~\ref{jmax} is valid in the small $\lambda$ and large $F$ regime (however, $F<\omega_0$) where $\bar j$ decreases due to decreasing overlap between Stark states. 

The scaling breaks down when with decreasing $F$, $\bar j$ approaches  the maximum $\bar j_{\mathrm{max}}$.  In Fig.~\ref{fig4}(c) we present results for larger $\lambda\in[0.2,2.0]$ which enables us to observe the evolution of $\bar j$ vs. $F$ as the system evolves from the weak EP coupling ($\lambda<1$) towards the strong EP coupling ($\lambda>1$) regime. 
With increasing $\lambda$,  the position of  $\bar j_{\mathrm{max}}$ shifts towards  larger  values of $F$ while it decreases in its magnitude.
The main difference between the weak and strong EP coupling regime emerges due to increasing energy gap $\Delta$ in the polaron excitation spectrum  \cite{osor1,osor2,vidmar10a} that for $\lambda>>1$ approaches $\Delta \sim \omega_0$. Due to large $\Delta$ at large $\lambda>1$, $\bar j$ remains zero until $F\sim F_{\mathrm{th}}$.

We summarize  the overview of numerical results with a  diagram describing different regimes characterized by distinct  short-time behaviors (after switching on $F$), presented in Fig.~\ref{fig4}(d).  
We distinguish four  different regimes: 
(i) The regime of damped free particle BO (DBO) for small values of $g$. 
(ii) Almost critically damped (CD) regime where steady state current is reached in a time shorter than or of the order of $t_B$, and the oscillations in the current are still visible, however $j(t)>0$ for any $t>0$.  
(iii) Polaron BO regime (PBO) where system evolves nearly adiabatically. Polaron Bloch oscillates within the polaron band and damping is exponentially small (numerically undetectable). 
In PBO average current remains zero and total energy remains periodic  within numerical accuracy and up to the largest measured time $t \leq 20t_B$.
(iv) Damped polaron BO regime (DPBO) where remnants of PBO are seen in $j(t)$ while there exists a measurable average current $\bar j>0$ within $t\leq 20t_B$.


\section{Conclusions}

{\it In summary} we list our main results. Using a time dependent Lanczos method we have followed the time evolution of the polaron  from its ground state towards  the  steady state after the electric field has been switched on. Different sizes of VHS have been used to ensure that presented results are valid in the thermodynamic limit.
Steady state conditions have been reached at intermediate to high electric fields and the current vs. voltage characteristics has been plotted for different regimes of EP couplings.
By calculating the electron-phonon
correlation function representing the time evolution of the polaron,
we show that the absorbed energy in the steady state
is deposited as an increasing number of phonon excitations
arranged as a growing tail behind the moving polaron.

The damped BO can be observed  in the extremely weak EP coupling limit. 
In the former case, period of BO $t_B=2\pi/F$ should be less than
the relaxation time $t_0/g^2$ related with the emission of phonons. 
A  large gap in the spectrum in the strong coupling regime is responsible for observation of nearly  perfect BO arising from the polaron motion within  the polaron band.
The breakdown of this quasiadiabatic regime  qualitatively 
resembles the Landau-Zener transition from the polaron band to higher excited states.
Analytical estimate for the  steady state current on the electric field and EP coupling constant at large fields is proposed and numerically tested. The unusual  steady state  current vs.  electric field  dependence, $\bar j\propto \sqrt{\lambda/F}$, valid at large $F$ and small $\lambda$, reflects the significance of coherent processes for proper description of polaron motion. In contrast, approaches calculating the steady state current  relying on probabilities for transitions between neighboring Wannier-Stark states, mediated by the EP coupling, yield  $\bar j\propto {\lambda/F}$\cite{emin,rott}.

\acknowledgments 

We acknowledge stimulating discussions with C.D. Batista and financial support of the SRA under grant P1-0044. J.B. and L.V. acknowledge financial support of the REIMEI project, JAEA, Japan. 

\appendix*

\section{Coherent propagation between Stark states}

To analyze the propagation of the polaron an alternative approach to driven Hamiltonian, 
Eq.(\ref{hol}), is to study eigenstates in a constant external electric field, i.e., the electron Hamiltonian is written as
\begin{equation}
H_e= -t_0 \sum_i (c^\dagger_{i+1} c_i +\mathrm{H.c.}) - F \sum_i i n_i,  \label{he}
\end{equation}
Since eigenstates of $H_e$ are localized Stark states we perform the transformation to new orthogonal basis,
\begin{equation}
\alpha_l = \sum_i w_{i-l} c_i,
\label{stark}
\end{equation}
where wavefunctions $w_j$ (being real) are localized in the interval $-L_S/2<j<L_S/2$ with $L_s \sim 4 t_0/F$,
and eigenenergies $\epsilon_l = F l +\epsilon_0$. 
In order to keep constant energy, the particle can propagate along the chain only by emitting (absorbing) phonons. The novel unperturbed term
\begin{equation}
H_0 = \sum _l \epsilon_l \alpha^\dagger_l \alpha_l + \omega_0 \sum_i a^\dagger_i a_i, 
\label{h0n}
\end{equation}
connects the average displacement in the Stark basis $\Delta \epsilon_l=F \Delta l$ to 
phonon generation  $\omega_0 \Delta N_{ph}$. 

In the following we consider only the simple commensurate case 
where $\omega_0/F = K $ is integer ($K>1$, i.e. $F<\omega_0$), where electrons can perform coherent hopping
between Stark states with $\Delta l =K$ keeping $E_0$ constant by emitting (or absorbing) single phonon via 
the coupling term $H' = g\sum_i n_i (a^\dagger_i + a_i)$. 
Also restricting phonon frequencies to $\omega_0<W=4t_0$ we remain in the regime $ K < L_S$.
We now construct the basis of possible coherent states having the same $E_0 \sim 0$
starting with a bare electron state at $l=0$ and generating novel states by application of $H'$
\begin{eqnarray}
|\psi _0 \rangle& =& \alpha_0^\dagger | 0 \rangle , \nonumber \\
|\psi _1^{j_1} \rangle& =& \alpha_K^\dagger a^\dagger_{j_1} | 0 \rangle , ...\nonumber \\
|\psi _m^{j_1 j_2 \dots j_m} \rangle& =& \alpha_{mK}^\dagger a^\dagger_{j_1} 
a^\dagger_{j_1} \dots a^\dagger_{j_m}  | 0 \rangle , 
\end{eqnarray}
whereby $j_m$ denote location of phonons. The matrix elements between subsequent
states can be evaluated explicitly by neglecting multiple occupations of sites 
(being rare for $K \gg 1$), i.e. $j_1 \neq j_2 \dots$
or equivalently simplifying boson factor for multiply occupied sites, i.e.,
\begin{eqnarray}
&&\langle \psi _m^{j_1 \dots j_{m-1} j_m} |H'| \psi _{m-1}^{j_1 \dots j_{m-1} } \rangle 
\nonumber \\
&&\sim g w_{j_m - mK} w_{j_m - (m-1)K} = T_{j_m}, \label{matr}
\end{eqnarray}
which depends within such an approximation only on $\tilde j_m = j_m- mK$.
We search now for the eigenstates in such a restricted space in the form
\begin{eqnarray}
|\Psi \rangle &=& b_0 |\psi _0 \rangle + \sum_{j_1} b_1^{j_1} |\psi _1^{j_1} \rangle + \dots
\nonumber\\ &+& \sum_{j_1 j_2 \dots j_m} b_m^{j_1j_2\dots j_m}  
|\psi _m^{j_1 j_2 \dots j_m} \rangle \dots, \label{psi}
\end{eqnarray}
where the energies $\tilde E$ are obtained solving the system
\begin{eqnarray}
\tilde E b_0 &=& \sum_{j_1} T_{j_1} b_1^{j_1}, \nonumber \\
\tilde E b_1^{j_1} &=& T_{j_1} b_0 + \sum_{ j_2} T_{j_2} b_2^{j_1j_2}, \label{ensys} \\
\tilde E b_m^{j_1\dots j_m} &=& T_{j_m} b_{m-1}^{j_1\dots j_{m-1}} 
+ \sum_{j_{m+1}} T_{j_{m+1} } b_{m+1}^{j_1\dots j_{m+1}},   \nonumber
\end{eqnarray}
With inserting the solutions of the Stark problem into Eq.(\ref{matr}), the branching system,
Eq.(\ref{ensys}), can be solved quite generally. Here, we are interested only in
a qualitative behavior, hence we use the simplification
\begin{equation}
T_j \sim \frac{g}{L_S} (-1)^{r_j}, \qquad -L_S/2 + K < j < L_S /2,
\label{simpli}
\end{equation}
and $T_j=0$ elsewhere, where the phase $(-1)^{r_j}$ emerges 
from fast varying Stark functions $w_j$ in Eq.(\ref{matr}).

Solutions of Eqs.(\ref{ensys}),(\ref{simpli}) can be found by an Ansatz
\begin{equation}
b_{m+1}^{j_1\dots j_{m+1}} \sim e^{-i p K} \frac{1}{\sqrt{\tilde L_S}} (-1)^{r_{j_{m+1} }}
b_m^{j_1\dots j_m} , \label{bb}
\end{equation}
where $\tilde L_S=L_S-K$.
The corresponding eigenenergies are
\begin{equation}
\tilde E = \tilde E_{p} = \frac{2g}{L_s} \sqrt{\tilde L_S} \cos (pK), \label{enp}
\end{equation}
which leads to group velocities in the tight-binding 
form $v_p \propto v_p = v_0 \sin(pK)$ with the maximum
\begin{equation}
v_0 \simeq \frac{2gK}{L_S} \sqrt{\tilde L_S} \sim  \frac{2gK}{\sqrt {L_S}} 
= \frac{g \omega_0}{\sqrt{ t_0 F}} . \label{v0}
\end{equation}

The derivation can be made more rigorous taking into account the actual Stark 
wavefunctions  $w_j$ and matrix elements Eq.(\ref{matr}). Still it is not expected to
change qualitatively the scaling of coherent group velocities $v_p$ with the 
maximum $v_0$, Eq.(\ref{v0}). It should be, however, reminded that 
we did not yet match the actual solution Eq.(\ref{enp}) with the boundary condition,
as determined, e.g.,  with the first equation in the system Eq.(\ref{ensys}).
Anyhow, it is expected that an eigenstate of stationary Hamiltonian,
as in Eq.(\ref{he}), cannot posess a finite steady current (solution being a 
superposition of $ \pm p$  eigenstates).  On the other hand, the driven system
and the time-dependent model, Eq.(\ref{ham}), clearly can generate the current 
$j(t)$ and in this sense induce solutions with the steady current $\bar j \propto 
j_0 = v_0$ following Eq.(\ref{v0}). Evidently, more rigorous relation between the 
eigenstates of the stationary case and the driven problem  is still desired.

\bibliography{manuphE}

\end{document}